\documentclass[iop,natbib209]{emulateapj}

\usepackage[usenames,dvipsnames,svgnames,table]{xcolor}

\usepackage{graphicx}
\usepackage{amsmath}
\usepackage{natbib}

\shorttitle{}
\shortauthors{Hsieh et al.}

\begin{document}

\title{The  Nuclear Filaments inside the Circumnuclear Disk in the Central 0.5 pc of the Galactic center}

\author{
        Pei-Ying Hsieh \altaffilmark{1},
        Patrick M. Koch\altaffilmark{1},
        Woong-Tae Kim \altaffilmark{2},
        Paul T. P. Ho\altaffilmark{1,3},
        Hsi-Wei Yen \altaffilmark{1},
        Nanase Harada\altaffilmark{1},
        Ya-Wen Tang\altaffilmark{1}        
\\pyhsieh@asiaa.sinica.edu.tw}

\affil{$^1$ Academia Sinica Institute of Astronomy and
       Astrophysics, P.O. Box 23-141, Taipei 10617, Taiwan, R.O.C.}
\affil{$^2$ Department of Physics \& Astronomy, Seoul National University, Seoul 151-742, Korea}
\affil{$^3$ East Asian Observatory, 660 N. Aohoku Place, University Park, Hilo, Hawaii 96720, U.S.A.}

\begin{abstract}

We present CS(7-6) line maps toward the central parsec of the Galactic Center (GC), conducted with the Atacama Large Millimeter/submillimeter Array (ALMA). The primary goal is to find and characterize the gas structure in the inner cavity of the circumnuclear disk (CND) in high resolution (1.3$\arcsec$=0.05 pc). Our large field-of-view mosaic maps -- combining interferometric and single-dish data that recover extended emission -- provide a first homogeneous look to resolve and link the molecular streamers in the CND with the neutral nuclear filaments newly detected within the central cavity of the CND. We find that the nuclear filaments are rotating with Keplerian velocities in a nearly face-on orbit with an inclination angle of $\sim10\degr-20\degr$ (radius $\le$ 0.5 pc).  This is in contrast to the CND which is highly inclined at $\sim65\degr-80\degr$ (radius $\sim$2-5 pc). Our analysis suggests a highly warped structure from the CND to the nuclear filaments. This result may hint that the nuclear filaments and the CND were created by different external clouds passing by Sgr A*.

\end{abstract}
\keywords{Galaxy: center --- radio lines: ISM --- ISM: molecules --- Galaxy: structure --- techniques: image processing}

\section{Introduction}

The number of detections of circumnuclear gas in the central parsecs of supermassive black holes (SMBHs) has increased over the recent years. Studies typically find a complex morphology with streamers, spirals, and warped rings/disks \citep{onishi15,neumayer07,tristram09,espada17,imanishi18,izumi18}.
Very often these nuclear structures are dynamically decoupled from the large-scale structures.
The clearest detections, due to its proximity, are in the molecular torus around the Galactic Center (GC), called  the circumnuclear disk/ring (CND or CNR) \citep[e.g.,][]{guesten87,serabyn89,lau,maria,martin12,liu12,great,mills13b,hsieh17,tsuboi18}.
The CND, with a radius from 2 to 5 pc, is rotating with respect to the SMBH, Srg A*. Embedded within the CND are the ionized gas streamers called SgrA West (mini-spiral) \citep[e.g.,][]{zhao09}. Being the closest molecular reservoir in the GC, the CND is critical in the understanding of the feeding of the nucleus \citep{ho91}.

One possible scenario for the formation of the CND involves tidal stripping.  In such a scenario a nearby molecular cloud within the Roche radius of Sgr A* is stripped and moves inward to settle down into the CND due to partial conservation of angular momentum \citep{sanders98,vollmer02,mapelli16}. The ``tidal debris'' is observed as multiple streamers in the dense molecular gas \citep{liu12,shunya17,hsieh17}. These streamers show a signature of rotation and inward radial motion with progressively higher velocities as the gas approaches the CND and finally end up corotating with the CND. These results from our team might suggest a possible schematic of the formation of the CND \citep{hsieh17,hsieh18}.
With the high angular resolution and sensitivity of ALMA, we are able to ask the next question: ''Is the molecular inflow able to be channeled past the CND?''
Using  early  ALMA  observations, \citet{moser17} presented the detection of molecular gas within 0.5 pc of Sgr A* (called the central association in their paper), which is physically located in the inner cavity of the CND with a higher excitation condition than the CND \citep{herrnstein02}.
This central association is proposed to link Sgr A* with the northeastern part of the CND, or the OH streamer southeast of the CND \citep{karlsson15}. The analysis in \citet{moser17} is limited by the small field-of-view of the data and the absence of the ALMA total power array.
In order to probe whether the infall process is ongoing from the CND to well within the central pc of Sgr A*, we present in this paper mosaic maps over the central 3$\arcmin$ by utilizing all possible UV sampling with ALMA. By using a high-density tracer, we are able to follow the kinematics of this central region all the way starting from the CND.

\section{ALMA Observations and Data Reduction}\label{sect-obs}

\begin{figure*}[bht!]
\begin{center}
\epsscale{0.6}
\includegraphics[scale=0.45,angle=0]{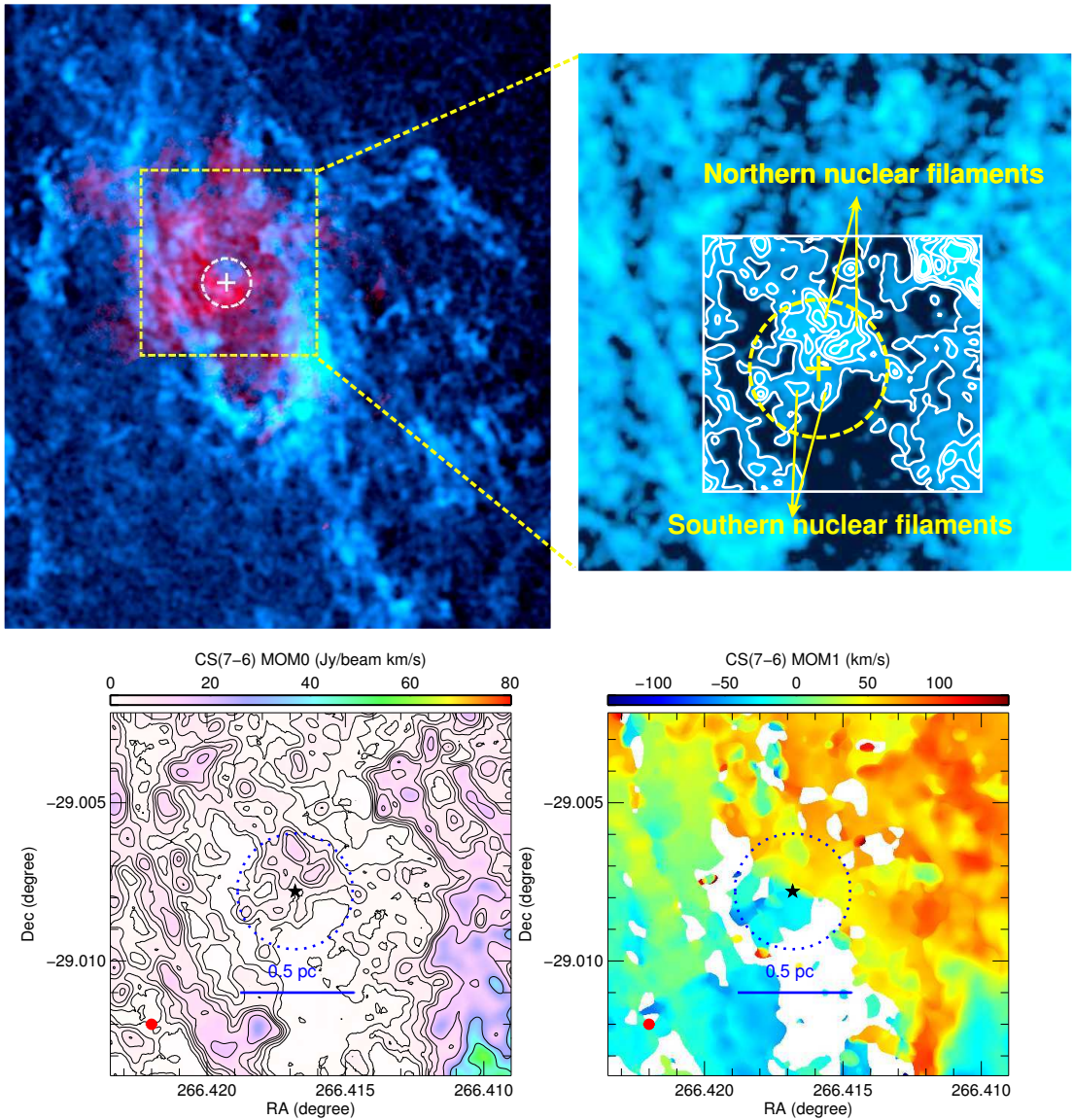}
\caption[]{\linespread{1}\selectfont{}
Upper: Color-composite image of ALMA CS(7-6) integrated intensity map tracing the inner region of the CND (blue) and HST Pa$\alpha$ map tracing the Sgr A West (red) \citep{dong11}. Left panel:  entire region observed with ALMA. Right panel:  zoomed-in image with contours of 0.16, 0.7, 1.2, 2.7, 6.5, 36.6 Jy beam$^{-1}$.
The location of Sgr A* is labeled with a cross. The circle shows a diameter of 0.5 pc. The nuclear filaments are located within this  circle. The resolution shown in the map is the original resolution of 0.8$\arcsec$ for the CS(7-6) data. Lower left panel: Integrated intensity map (MOM0)  of CS(7-6). The CS(7-6) contours are  1.5, 3, 5, 7, 9, 15, 30, 45, 60 Jy beam$^{-1}$ km s$^{-1}$.  Lower right panel: intensity-weighted velocity map (MOM1) of CS(7-6). A smoothed beam of 1.3$\arcsec$ ($=$ 0.05 pc), shown as red dot, is used to enhance the extended emission connecting the CND and the nuclear filaments.
}
\label{fig-m0}
\end{center}
\end{figure*}

ALMA observations toward the central 3$\arcmin$ of the GC were carried out using 43 12-m antennas (project code: 2017.1.00040.S) and the total-power array. 
We observed several lines and transitions among which the results for CS (7-6) ($f_{\rm rest}=$342.882 GHz) are reported in this paper.
The default channel width of the spectral windows was 244.141 kHz. The observations consisted of four 150-pointing mosaic maps to cover the central 3$\arcmin$ region.
Calibration of the raw visibilities was performed with the pipeline and manual reduction script for the cycle-5 data in Common Astronomy Software Applications (CASA v5.1.1-5). To recover the zero-spacing in the interferometric data, we combined the 12 m, 7 m, and single-dish data. The CASA tasks Tclean and Feather were used to deconvolve and merge the interferometric and single-dish maps. The task Mstrasform was used to subtract the continuum emission. We processed the four individual mosaics separately, and combined them into one map.
We used the Briggs robust parameter of 0.5. The image cubes were made at a velocity resolution of 2 km s$^{-1}$ to enhance the signal-to-noise ratio.
The native resolution of the CS(7-6) map is 0.8$\arcsec$. To reveal the extended emission in the inner edge of the CND, we smoothed the maps to 1.3$\arcsec$ ($=0.05$ pc). The noise level is 11 mJy of the smoothed map. In this paper, we  present the analysis on the central 5 pc of the GC. 
The results from the additional lines and transitions will be presented in a forthcoming paper.  

\section{Results}

\begin{figure*}[bht]
\begin{center}
\epsscale{0.6}
\includegraphics[scale=0.7,angle=-90]{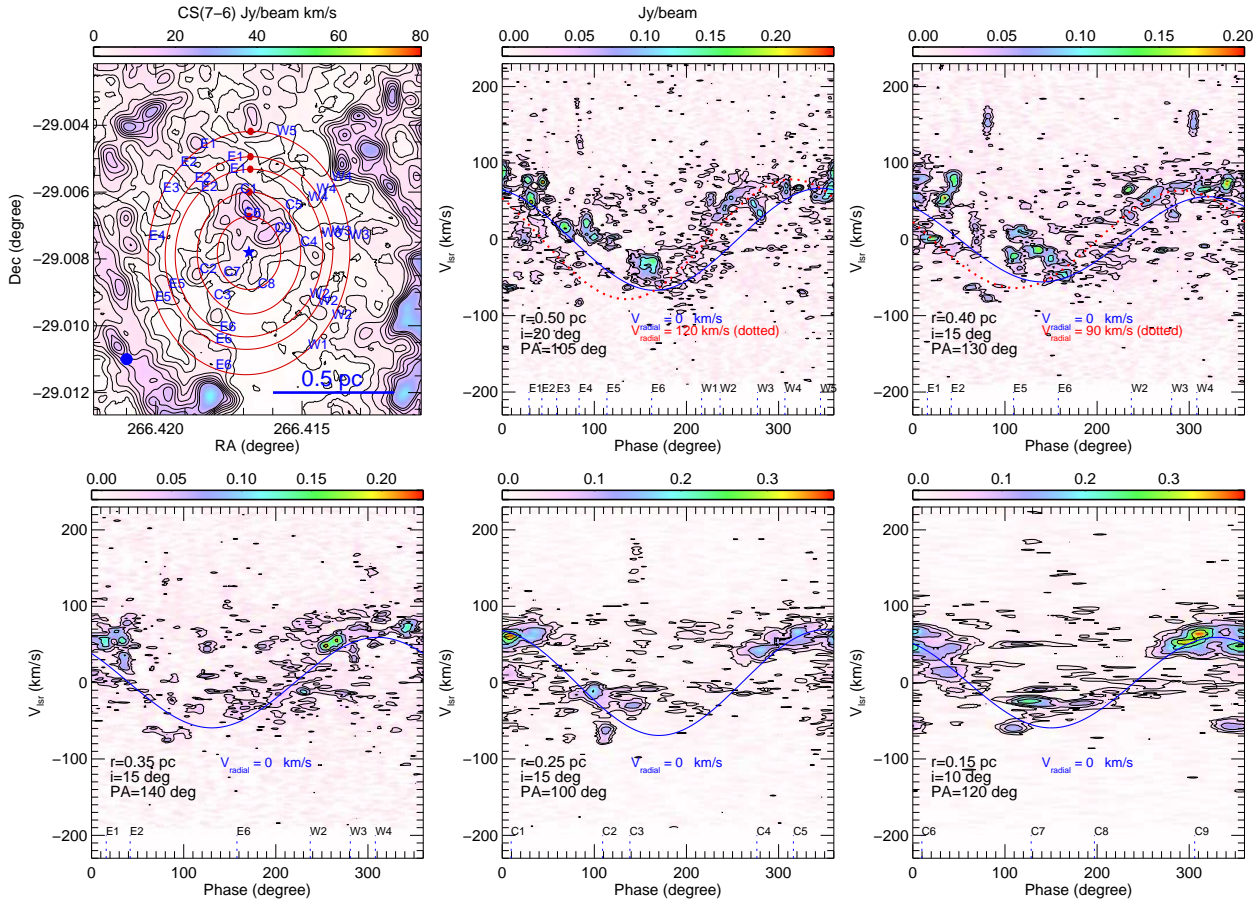}
\caption[]{\linespread{1}\selectfont{}
Line-of-sight velocities as a function of the phase angle along ellipses in the upper left panel at intrinsic radii of 0.5, 0.4, 0.35, 0.25, and 0.15 pc, in comparison with best-fit Keplerian velocities (blue lines). The contours are 0.025, 0.05, 0.1, 0.2, 0.3, 0.5, 0.7 Jy beam$^{-1}$.
 The corresponding orbital parameters are labeled in the lower left corners of the phase plots. The locations of E1...E6, W1...W5, and C1...C9 are labeled to mark the corresponding features in the intensity map and the phase plots. At radii of 0.4 and 0.5 pc, an additional radial inward velocity is required to fit the western part of the nuclear filaments (red dotted curve). The velocity structure at the other radii can be adequately described by a pure Keplerian motion. The starting positions ($0\degr$) of the trajectories are labeled by the red dots, and the phase is counterclockwise from the north. The $PA$ is defined starting from east clockwise.
}
\label{fig-vgb}
\end{center}
\end{figure*}

Figure~\ref{fig-m0} displays the integrated CS(7-6) intensity maps (MOM0) in the inner region of the CND. The detection of molecular gas within a projected diameter of 0.5 pc around Sgr A* is called the central association in \citet{moser17}.
This central association extends mostly along a north-south orientation, centered on Sgr A*.  This extended molecular gas/emission appears to consist of a few clumps or filaments. Because of their filamentary morphology apparently connecting to the CND, we call them {\it  nuclear filaments} in this paper.
The emission from the northern  nuclear filaments is stronger than from the southern  nuclear filaments. The lengths and widths of these filamentary structures  are $\sim0.2$ pc $\times0.1$ pc. In general, the CS(7-6) line morphology is similar to the CS(5-4) in \citet{moser17}. However, our wide-field map combining interferometric and single-dish data recovers the extended emission, especially in the inner edge of the CND. This newly detected structural connection of the CND with the nuclear
filaments appears clearer and more robust than in previous maps \citep{moser17,tsuboi18}. Moreover, the high-excitation CS(7-6) line also better discriminates against lower-excitation material and reveals the nuclear filaments more clearly than in CO(3-2) \citep{javier18}. In order to enhance the extended emission, we smooth the CS(7-6) data to 1.3$\arcsec$.  In what follows, we analyze the smoothed data. 

In Figure~\ref{fig-m0} we also show the intensity-weighted velocity map (MOM1) of CS(7-6). Although the kinematics of the sub-structures in this region are complicated, the CND and the nuclear filaments are overall dominated by rotation, from blueshifted (southeast) to redshifted (northwest), which indicates that the  nuclear filaments are  rotating about Sgr A* in the same sense as the CND \footnote{We note that the regions filled with nuclear filaments are still rather small. Thus, dynamical shear (induced possibly by tidal forces) and clumps having different (non-rotating) velocities can be also be present.}.

\begin{figure}[bht]
\begin{center}
\epsscale{0.6}
\includegraphics[scale=0.9,angle=0]{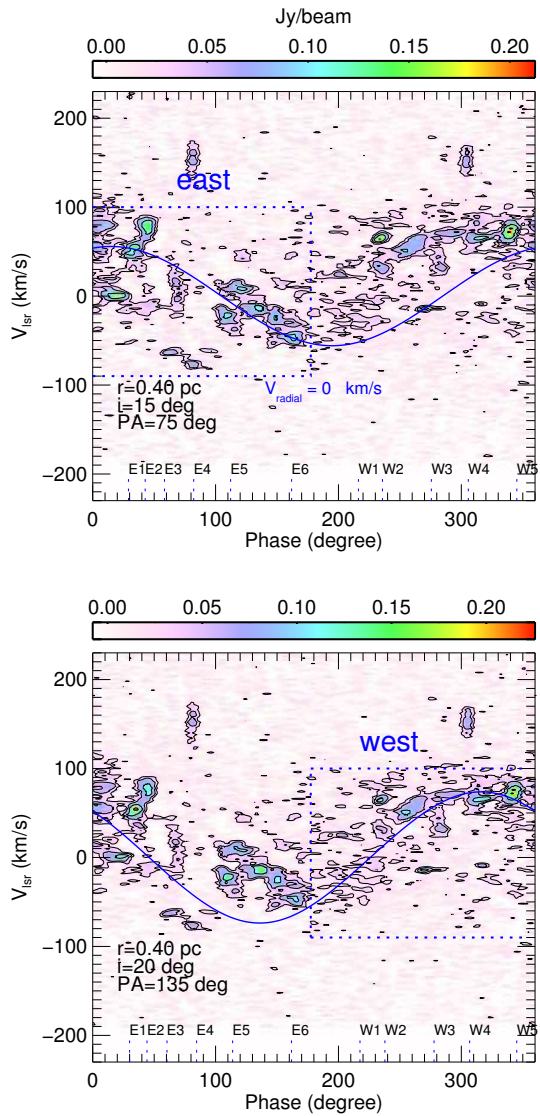}
\caption[]{\linespread{1}\selectfont{}
Line-of-sight velocity along the ellipse in the upper left panel of Figure 2 at an intrinsic radius of 0.4 pc.  Indicated are the best-fit  parameters for a Keplerian rotation (no radial inward velocity).
Here, the eastern and western nuclear filaments are fitted separately by a purely Keplerian model with different inclination and position angle: $i=15^{\circ}$ and PA=$75^{\circ}$ for the eastern nuclear filaments in the upper panel, and  $i=20^{\circ}$ and PA=$135^{\circ}$ for the western nuclear filaments in the lower panel. The fitting regions are marked by the blue rectangular box in each panel.}
\label{fig-vgb1}
\end{center}
\end{figure}

Since the MOM1 map indicates a rotating motion of the nuclear filaments, for a more quantitative analysis, we generate Keplerian rotation curves, and overlay them on the phase plots with the CS(7-6) data in Figure~\ref{fig-vgb}.
Our goals here are  to investigate whether the nuclear filaments are gravitationally bound by the central mass concentration and whether they originate from the CND. The CND can be described by three distinct rotating structures with circular orbits \citep{martin12}.
Therefore, we start this analysis with the simplest rotating ring model \citep[e.g.][]{guesten87,jackson93,martin12}. The enclosed mass $M(r)$ is adopted from the literature \citep[see eq. 5 in][]{genzel10} to account for the total mass of the enclosed stars and Sgr A* \citep[mass of Sgr A*$=4\times10^{6}~M_{\odot}$;][]{ghez08}.

To probe the kinematic structures as a function of distance to Sgr A*, we extract data centered on Sgr A* at five intrinsic radii of 0.5 pc, 0.4 pc, 0.35 pc, 0.25 pc, and 0.15 pc.
The data points at the different radii for fitting are chosen with a separation of one beam size, and can thus be considered as being independent.
The free parameters with their fitting ranges are the inclination (5$\degr$-80$\degr$ in steps of 5$\degr$), position angle (70$\degr$-150$\degr$ in steps of 5$\degr$), and the radial velocity (0 km s$^{-1}$-120 km s$^{-1}$ in steps of 10 km s$^{-1}$). The best-fit circular orbits
of the line-of-sight velocity $V_{\rm los}$ determined by the mininum $\chi^2$ (red ellipses after projection of the best-fit inclination and position angles) and phase plots are presented in Figure~\ref{fig-vgb}. Overall, the nuclear filaments can be described by Keplerian rotation, which may extend from a radius of 0.5 pc, i.e.,  within the inner edge of the CND (radius of $\sim$1 pc) down to the nuclear filaments.
Nevertheless, we notice that a single Keplerian rotation is unable to fit the centroids of  all features simultaneously at the radii of 0.4 and 0.5 pc. The deviation from Keplerian rotation is likely caused  by the inward radial velocities (90-120 km s$^{-1}$) of E1-E3 and W1-W3.  Alternatively, Figure~\ref{fig-vgb1} shows that the eastern and western sides at a radius of 0.4 pc can be described by pure Keplerian motions with different orbital parameters.

\begin{figure}[bht]
\begin{center}
\epsscale{0.45}
\includegraphics[scale=0.2,angle=0]{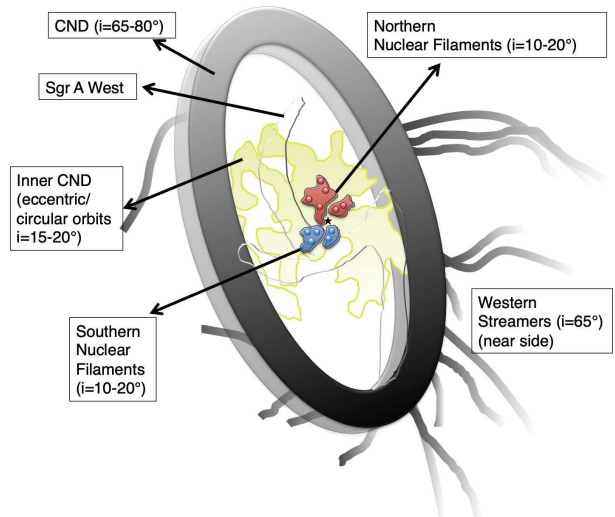}
\caption[]{\linespread{1}\selectfont{}
Schematic of the CND, Sgr A West, the nuclear filaments, the western streamers, and the extended molecular gas (in light green) between the CND and the nuclear filaments. The red and blue colors indicate redshifted and blueshifted velocities of the northern and southern nuclear filaments with respect to Sgr A* marked by the black star, respectively. Our data suggest that the nuclear filaments and the emission within the inner CND have a face-on geometry, but the CND and western streamers are known to have higher inclinations.
}
\label{fig-model}
\end{center}
\end{figure}

Our analysis suggests that the inclinations vary as a function of radius from 20$\degr$ to 10$\degr$. A higher inclination (65$\degr$ to 80$\degr$) is not able to explain the gross feature within the radius of 0.5 pc. This implies that the nuclear filaments (and the inner cavity of the CND) could have a face-on geometry with intrinsic velocities $\ge300$ km s$^{-1}$ at a (de-projected) radius of 0.15 pc. This range of the inclination angle  of the nuclear filaments is quite  different from that of the CND \citep[65$\degr$ to 80$\degr$;][]{martin12,hsieh17,javier18} if the assumption of Keplerian motion holds.
Figure~\ref{fig-model} illustrates the nuclear structures including the CND, the nuclear filaments, and the extended molecular gas  that we detected.

\section{Discussion}

Our ALMA observations successfully detect and resolve the nuclear gas within 0.5 pc of Sgr A*. With our mosaic map combined with single dish data, we find that the nuclear filaments rotate in the same sense as the CND with respect to Sgr A*. However, unlike the CND, which is largely tilted with an inclination angle between $65\degr$ to $80\degr$, the spatially resolved nuclear filaments in our ALMA map indicate a low inclination angle between 10$\degr$ and 20$\degr$, if they follow Keplerian rotation. However, we find that at radii of 0.4 and 0.5 pc not all the features in the inner edge of the CND are in pure Keplerian rotation. The deviation from circular rotation can be explained by an additional inward radial motion (90--120 km s$^{-1}$). This radial motion may occur at several eccentric orbits in this crowded region, similarly  to the mini-spiral shown in \citet{zhao09}. A possible constraint on eccentricity can be estimated by the ratio of tangential ($V_{\tan}$) and radial velocities ($V_{\rm rad}$) of an eccentric Keplerian orbit,
\begin{equation*}\label{eq1}
\frac{V_{\rm rad}}{V_{\rm tan}} = \frac{e\sin(\theta)}{1+e\cos(\theta)},
\end{equation*}
where $e$ and $\theta$ are the eccentricity and the angle between the current location of the clouds and the periapsis of Sgr A*, respectively.
Using $V_{\rm rad}$=120 km s$^{-1}$ and $V_{\rm tan}$= 200 km s$^{-1}$ (which is a rotational velocity of  68 km s$^{-1}$ corrected for an  inclination of 20$^\circ$), fixing $\theta$ to a certain range of locations, we obtain $e \geq$ 0.6 and $\theta \geq 60^\circ$ (i.e., far from the periapsis of Sgr A*) to match the ${V_{\rm rad}}/{V_{\rm tan}}$ ratio. Similar results are obtained for the gas at $r=$0.4 and 0.5 pc.

The existence of multiple orbits seems very attractive because  it is possible to treat the western and eastern sides (at $r=0.4$ pc) as different groups of circular orbits  (Figure~\ref{fig-vgb1}), although it is nontrivial to constrain orbital parameters for both eccentric and circular orbits.  The variations of inclinations and position angles with radius suggest that nuclear filaments are progressively warped as they move inward from the CND  (Figure~\ref{fig-vgb}).
It is already shown that the motion of the CND itself requires a bundle of non-uniformly rotating streamers of slightly different inclinations \citep{martin12,javier18}. It, nevertheless, is still intriguing  that the nuclear filaments indeed are low-inclination structures and at the same time appear to extend to the inner cavity of the CND, including the ``triop'' cloud named by \citet{moser17}.

We do not see a clear connection between the features reported here and the previous reported high-velocity CO(3-2) clumps \citep{javier18}.
The CS counterparts of the CO(3-2) high-velocity clouds are marginally detected in CS(7-6) (e.g., C3 at a de-projected radius of 0.25 pc in Figure~\ref{fig-vgb}). This might suggest that the high-velocity clouds have densities lower than the nuclear filaments. Our CS(7-6) data detect relatively denser components in this region.

Here we postulate a few possible origins of the nuclear filaments: (1) tidal leftover of an infalling cloud from the CND \citep{moser17}; (2) accretion starting from the inward radial motion in the CND \citep{hsieh17}; (3) tidal leftover of nearby clouds similar to the 50 km s$^{-1}$ cloud and 20 km s$^{-1}$ cloud, formed at the same epoch as the CND or at different epochs. The scenario (1) seems to be the most natural way to create the nuclear filaments because the gas densities of the CND are lower than the Roche limit. If the assumption of Keplerian motion is valid, however, it is very difficult to form a system with inclination angles so different from the CND's over a short time, unless the infalling cloud has a size comparable to the size of the CND (the dynamical time scale of the nuclear filaments is 5600 years at a radius of 0.25 pc). For scenario (2), can the CND feed gas towards Sgr A* via the nuclear filaments?
The inward radial motion in the CND is about 23--50 km s$^{-1}$ \citep{hsieh17,tsuboi18}. To create the nuclear filaments with a mass of 5 $M_{\odot}$ \citep{moser17}, it takes about 7800 years to accrete from 0.55 to 0.15 pc. This  seems plausible because it is similar to the rotation time scale, and the clouds may have not yet been disrupted by the tidal force, although the slow accretion would likely preserve an inclination angle similar to the CND's. Furthermore, the previous estimates of radial inward motions are based on the assumption of circular orbits, and the possibility that the inward radial velocities are due to eccentric orbits cannot be excluded. An inflow velocity, amounting to a quarter of the  rotational velocity, is necessary to explain the observed magnetic field configuration near the CND \citep{hsieh18}. Measuring  radial inward motions would need  full orbital modeling in the future  \citep{paumard04,zhao09}.
For scenario (3), it is possible to obtain extended molecular gas structures within 0.5 pc if a cloud with homogeneous density falls in from 26 pc towards an SMBH on a nearly radial orbit with non-vanishing angular momentum \citep{mapelli16}. In such a scenario, the infalling cloud is quickly disrupted by the SMBH, and the tidal force stretches the cloud to form a group of streamers within $2.5\times10^{5}$ yr. These streamers are then trapped by the SMBH and form two rings with radii of $\sim$0.4 pc and $\sim$2 pc, although the ring sizes and formation time may depend on the chosen parameters.

Yet an alternative interpretation is that the nuclear filaments have an inclination similar to the CND's,  but  their low projected velocities are sub-Keplerian due to radiation and thermal pressure.  For a comparison, the molecular torus in the nuclear region of NGC 1068 is interpreted as a sub-Keplerian rotation because the derived enclosed mass is 10 times smaller than the observed black hole mass under Keplerian motion. \citet{chan17} found that a sub-Keplerian torus is able and required to survive mass loss for multiple orbits in the presence of strong radiation pressure. If the assumption of  Keplerian motion is not valid, the nuclear filaments are possibly linked to and fed by the CND via gradual accretion or  a cloud originating from the CND.

\section{Summary} 

We present an analysis of the molecular nuclear filaments within the central 0.5 pc of the GC. The wide-field ALMA and single-dish combined map enables us for the first time to probe the dynamical picture of the infall scenario connecting from the CND to the inner nuclear filaments. With our ALMA CS(7-6) map, we find that the nuclear filaments are rotating in the same sense as the CND, but have a nearly face-on geometry with an inclination angle of $\sim10\degr-20\degr$. This is significantly different from the highly inclined CND. This result may suggest that the nuclear filaments and the CND were created during different passages of external clouds  passing by Sgr A*.

\acknowledgements

We are grateful to the referee for a thorough and insightful report which greatly helped to improve the paper.  P.-Y. H. is supported by the Ministry of Science and Technology (MoST) of Taiwan through the grants MOST 107-2119-M-001-041. P. M. K. acknowledges support from MOST 108-2112-M-001-012 and MOST 107-2119-M-001-023, and from an  Academia Sinica Career Development Award.  P. T.-P. H. acknowledges support from MOST 105-2112-M-001-025-MY3 and MOST 108-2112-M-001-016. H.-W. Y. acknowledges support from MOST 108-2112-M-001-003-MY2. Y.-W. T. acknowledges support from the MoST 108-2112-M-001-004-MY2. The work of W.-T. K. was supported by the National Research Foundation of Korea (NRF) grant funded by the Korea government (MSIT) (2019R1A2C1004857).


\end{document}